# Integrated InP-based transmitter for continuous-variable quantum key distribution

J. ALDAMA,[1,*] S. SARMIENTO,[1] L. TRIGO VIDARTE,[1] S. ETCHEVERRY,[1] I. LÓPEZ GRANDE,[1] L. CASTELVERO,[1] A. HINOJOSA,[2] T. BECKERWERTH,[3] Y. PIÉTRI,[4] A. RHOUNI,[4] E. DIAMANTI,[4] AND V. PRUNERI[1,5,6]

[1]*ICFO-Institut de Ciencies Fotoniques, The Barcelona Institute of Science and Technology, Castelldefels (Barcelona) 08860, Spain*
[2]*VLC Photonics S.L–Ed. 9B–D2, UPV, Camino de Vera s/n, 46022 Valencia, Spain*
[3]*Fraunhofer Heinrich Hertz Institute (HHI), Einsteinufer 37, 10587 Berlin, Germany*
[4]*Sorbonne Université, CNRS, LIP6, 4 Place Jussieu, F-75005 Paris, France*
[5]*ICREA-Institució Catalana de Recerca i Estudis Avançats, Barcelona, 08010, Spain*
[6]*valerio.pruneri@icfo.eu*
**jennifer.aldama@icfo.eu*

**Abstract:** Developing quantum key distribution (QKD) systems using monolithic photonic integrated circuits (PICs) can accelerate their adoption by a wide range of markets, thanks to the potential reduction in size, complexity of the overall system, power consumption, and production cost. In this work, we design, fabricate and characterize an InP-based PIC transmitter for continuous-variable (CV) QKD applications. In a proof-of-principle experiment implementing a pulsed Gaussian-modulated coherent state (GMCS) CV-QKD protocol over an optical fiber channel of 11 km, the system showed a performance compatible with a secret key rate of 78 kbps in the asymptotic regime. These results show the potential of InP technologies to integrate CV-QKD systems onto a monolithic platform.

## 1. Introduction

Current cryptographic protocols employed to safeguard digital data rely on highly complex computational methods. However, computational-based cryptography is susceptible to attacks performed by quantum computers, undetected eavesdropping, and "store-now-decrypt-later" attacks. To address these challenges, quantum cryptography provides a promising solution: quantum key distribution (QKD). QKD protocols allow two distant parties to share a cryptographic secret key through a public channel with information-theoretic security (ITS), preventing eavesdroppers from gaining access to the key without being detected [1,2].

Commercial QKD systems are already available, although they remain expensive, bulky, and relatively challenging to operate. Consequently, QKD is difficult to use on a large scale. To overcome these issues, the photonic integration of QKD systems can significantly reduce their cost, size, weight, and power consumption, and facilitate mass production, making QKD accessible to a wider range of users [3,4]. For photonic integrated QKD implementations, continuous-variable QKD (CV-QKD) is a suitable choice due to all components can potentially be integrated into the same chip, including coherent detectors. Moreover, CV-QKD is compatible with classical coherent optical communication networks (a prevailing standard in metro and long-haul optical networking) and can coexist with classical light dataflows in the same fiber [5]. A well-known protocol for the implementation of CV-QKD systems is Gaussian-modulated coherent-state (GMCS), where the quadrature components of coherent states are randomly modulated following a zero-center Gaussian distribution [6,7]. GMCS has been proven to be secure against the most general attacks, including in the finite-size setting [8–10].

Recently, various research groups have endeavored to downsize QKD systems using different photonic integrated circuit (PIC) platforms, such as silicon, III-V compound semiconductors, and lithium niobate (LiNbO3) [3,11–13]. Among these, several on-chip CV-QKD systems have been investigated by combining external lasers and integrated components on a silicon photonics platform [14–22]. Although silicon photonics offers numerous advantages, telecom lasers cannot be monolithically integrated, requiring non-standard hybrid integration procedures. In a recent work [23], low-linewidth lasers based on an InP reflective semiconductor optical amplifier (RSOA) butt-coupled to a low-loss silicon nitride cavity were used in a bulky CV-QKD system showing promising results. InP has therefore the inherent advantage of allowing the monolithic integration of a range of components necessary for CV-QKD systems, such as lasers, detectors, and high-speed modulators.

In this paper, we propose an InP-based PIC transmitter for CV-QKD implementing the pulsed GMCS protocol, extending and consolidating the results presented in [24]. In Section 2, we describe our PIC transmitter (TX) design and propose an experimental proof-of-principle CV-QKD setup to assess the potential performance of our design. Section 3 presents the electro-optical characterization of the most relevant building blocks of the PIC, as well as the CV-QKD experimental system-level results, including the secret key rate estimation in the asymptotic and finite size regimes. Finally, we draw the main conclusions and prospects of our work.

## 2. Description of the CV-QKD TX and experimental setup

### *2.1 Building blocks and PIC scheme*

Our focus is to monolithically integrate the building blocks required for the modulation stage of a CV-QKD system, assuming a continuous wave (CW) laser input, with the objective of providing an output signal compatible with CV-QKD. With this purpose, we designed a versatile and research-friendly system with redundant paths, that allow testing the different components individually (see Fig. 1(a)). The PIC system can be divided into three main building blocks: (1) an electro-absorption modulator (EAM), to carve the optical pulses; (2) an IQ modulator, to modulate the optical signal; and (3) a variable optical attenuator (VOA), to adapt the optical signal to CV-QKD levels.

Our proposed PIC TX includes active components such as an EAM, thermo-optic phase shifters (TOPSs), and current injection phase shifters (CIPSs) modulators. It also contains passive components like waveguides, spot-size converters, and 2x2 power splitters based on multimode interferometers (MMIs); see Fig. 1(a). The PIC was designed to operate at the C-band (~1550 nm) and fabricated using the available building blocks in the HHI's transceiver (TX-RX) platform for multi-project wafers (MPWs).

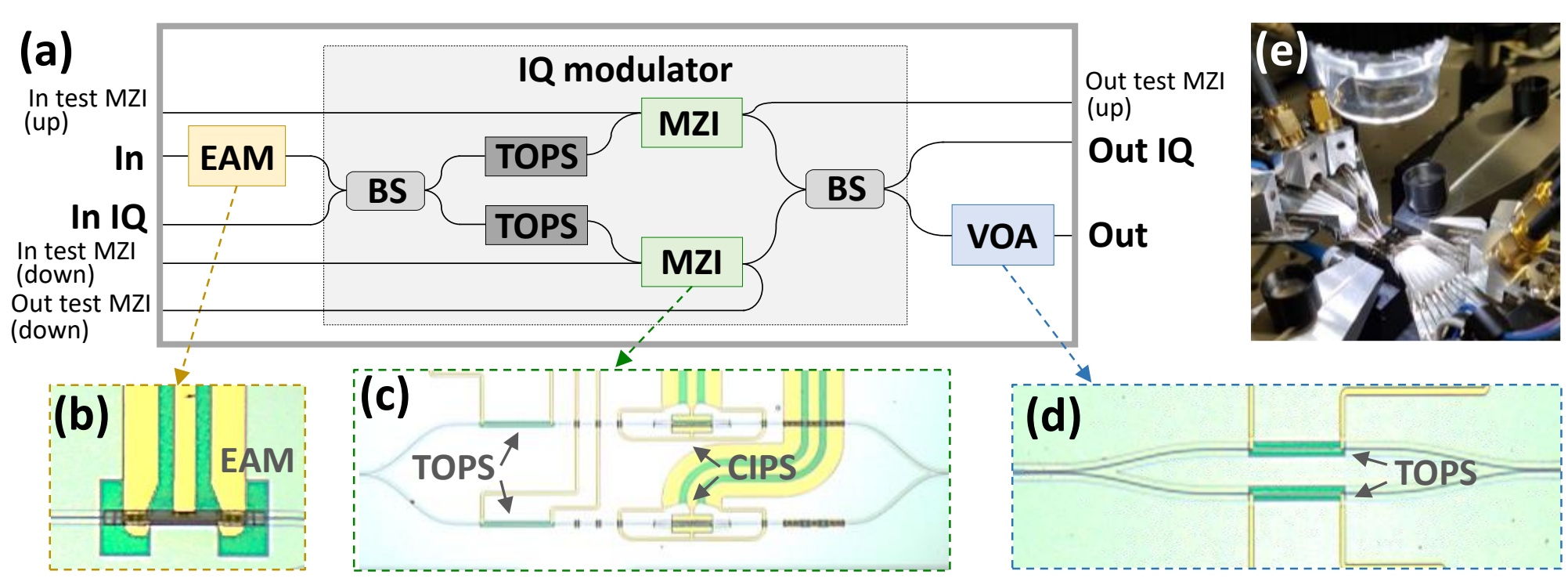

Fig. 1. (a) Block diagram of the CV-QKD PIC TX. EAM, electro-absorption modulator; BS, beam splitter 2x2 MMI 50:50 ratio; TOPS, thermo-optic phase shifter; MZI, Mach-Zehnder interferometer; VOA, variable optical attenuator. Microscope image of the (b) EAM; (c) MZI (CIPS, current-injection phase shifter); (d) VOA. (e) Setup including the fabricated $12 \times 6$ mm$^2$ InP-based CV-QKD PIC TX, Fraunhofer HHI Foundry, in the center.

The fabricated EAM is 200 µm long and its purpose is to pulse the light with a high extinction ratio (ER) (Fig. 1(b)). Preparation of the quantum states is carried out in the IQ modulator, which is composed of two nested Mach-Zehnder interferometers, MZIs (Fig. 1(c)). Each MZI combines two CIPSs and two TOPSs. The phase shift of $\pi/2$ between these two MZIs is set with two TOPSs. The VOA is also based on a MZI design, where the interferometric process is managed by two TOPSs (Fig. 1(d)). The purpose of the VOA is to attenuate the light power to the quantum level. Extra ports were added on the side of the main input edge (e.g., In IQ, In test MZI) and output port edge (e.g., Out IQ, Out test MZI) for monitoring the IQ modulator output as well as the performance of the VOA and MZIs separately.

For fiber coupling, our PIC TX includes spot size converters (SSCs) at the facets to increase the coupling efficiency to standard single-mode fiber (SMF). To avoid stray light issues, the optical ports were positioned facet to facet and shifted in our chip design. The electrical DC and RF components were placed along the longest edges of the chip while the optical ports were positioned along the shortest sides. The standard distance between the SSCs was set at 125 µm, while the electrical DC and ground-signal-ground (GSG) RF pads had a pitch of 150 and 130 µm, respectively. An image of the PIC mounted in our probe station can be seen in Fig. 1(e).

### 2.2 Experimental setup of the CV-QKD system

The proposed scheme can accommodate different CV-QKD implementations using CW and pulsed light, including real and transmitted local oscillator (LO) configurations, as well as flexible modulation schemes (e.g., Gaussian, discrete-modulation). To focus on the performance of the PIC, we chose a transmitted LO (TLO) scheme with Gaussian modulation using externally pulsed optical light (see Fig. 2 (a)) and heterodyne detection. The TLO scheme reduces complexity in phase noise compensation, as it circumvents the need for sophisticated carrier frequency recovery within the digital signal processing (DSP) sequence of the system [25,26]. This configuration allows testing the IQ modulator and VOA at system level. As we will see later, the elevated insertion losses (IL) in the EAM path ruled out the possibility of using it to carve the pulses in practice, but it was still possible to characterize its individual performance.

In our system, the light source, shared by Alice and Bob, is a Pure Photonics continuous-wave (CW) external cavity laser with a 10 kHz nominal linewidth and biased to emit 50 mW of optical power at the frequency of 193.4 THz (1550 nm wavelength). The CW laser, located at Bob's side, is followed by a 90:10 beam splitter (BS), where, for simplicity, the 90% BS output is directly used as Bob's CW local oscillator (LO), and the rest of the light is sent to Alice, where it is carved into pulses by an external lithium niobate amplitude modulator (AM). The AM is driven by a signal coming from a field-programmable gate array (FPGA) electronic board with 1 GSa/s 16-bit nominal resolution digital-to-analog converter unit, properly amplified electronically to reach the switching voltages required by the AM. The pulsed light is edge-coupled into the PIC after optimizing the polarization with a manual polarization controller (PC).

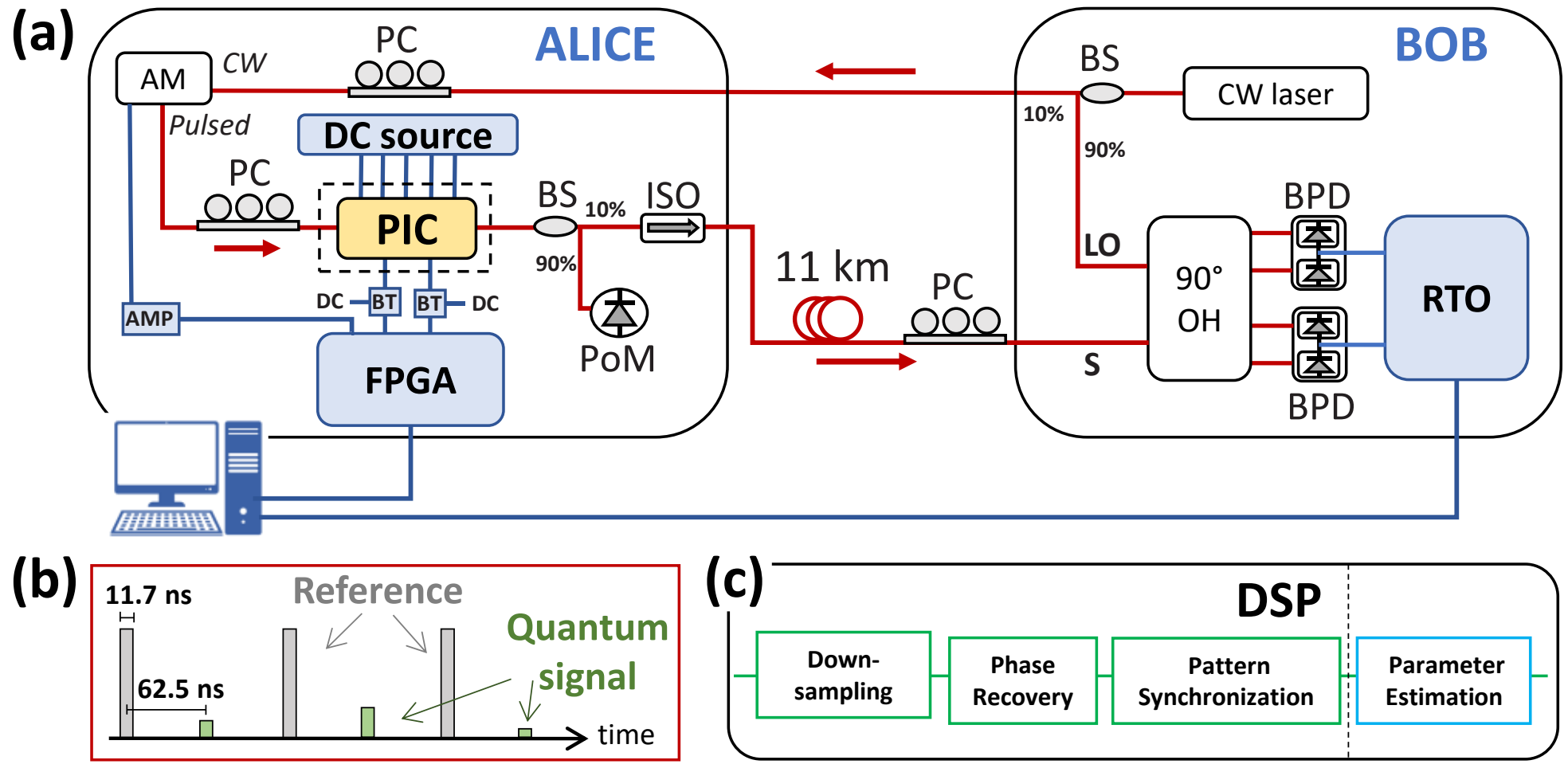

Fig. 2. (a) CV-QKD system. S, signal; LO, local oscillator; BS, beam splitter; PC, polarization controller; AM, amplitude modulator; AMP, electrical amplifier; PIC, photonic integrated circuit; BT, bias tee; FPGA, field-programmable gate array; PoM, power meter; ISO, isolator; OH, optical hybrid; BPD, balanced photodetector; RTO, real-time oscilloscope. (b) Pulses obtained at the output of the PIC and sent by Alice to Bob. Reference pulses were interleaved with pulses containing the quantum signal. (c) Digital signal processing (DSP) chain performed to the analysis of the data and parameter estimation.

In the PIC, the DC signals applied to the TOPSs are regulated with a NICSLAB DC multichannel source. The RF signals applied to the IQ modulator are generated by the FPGA in combination with a bias-tee (BT), where the DC input is used to compensate the threshold current, and the RF input is used to apply the modulated signal. In this proof-of-principle implementation, the Gaussian modulated symbols consist of two independent sets of 2040 pseudo-random values sent in cycles to the IQ modulator of the PIC. The quantum symbols modulated according to zero-centered Gaussian random distributions were interleaved in time with reference pulses of constant amplitude and alternating signs to facilitate the accuracy of the phase recovery at Bob's site (Fig. 2(b)). The external AM sets the initial amplitude difference between quantum and reference pulses. The repetition rate of the generated pulses is 16 MHz with a pulse width of 11.7 ns. Afterward, the coherent states of light at the PIC output are sent to a 90:10 beam splitter (BS). 90% of the light is measured using the power meter (PoM) to estimate the modulation variance ($V_A$), which is related to the mean photon number $\langle n \rangle$ at Alice's output by the formula $V_A = 2\langle n \rangle$. The remaining 10% of the light is sent to Bob through an 11 km fiber spool (Corning SMF-28 ULL fiber), with measured optical losses of 2.04 dB.

At Bob's optical signal channel input, the PC is used to maximize the interference visibility by adjusting the polarization of the signal with that of the LO. The signal sent by Alice interferes with the LO in a COH24 Kylia 90º optical hybrid (90° OH). The outputs of the OH are then detected using two 500 MHz bandwidth FEMTO balanced photodetectors (BPDs), and digitized with a 2 GSa/s real-time oscilloscope (RTO) with a digital filter with 50 MHz bandwidth. After that, the DSP is performed offline on a desktop computer following the 5 steps shown in Fig. 2(c) and explained in detail in Section 3.2. The secret key rate (SKR) estimates were calculated considering trusted receiver electronic noise, and reverse reconciliation [6,27]. Moreover, the SKR was evaluated for both the asymptotic limit and finite size regime.

## 3. Analysis and results

### 3.1 Block-by-block electro-optical characterization

Results of the steady-state electro-optical characterization of the main blocks (EAM, MZI, and VOA) are presented in Fig. 3 and Fig. 4, where the measured transmittance of each component is normalized to the maximum value of each plot. The measurements were performed using auxiliary paths connected to each block, a photodiode to measure the optical output from each block, and a source to perform sweeps in voltage and current [28].

The electro-optical results of the EAM for both TM and TE polarizations of the incident light are presented in Fig. 3(a) and Fig. 3(b), respectively. In both figures, changes in the EAM transmittance can be seen at different optical frequencies when a sweep in voltage is applied to the EAM. The highest ER value was $ER_{EAM}$=28.5 dB and was obtained at 191.55 THz and TE polarization (Fig. 3(b)). Meanwhile, the results using TM polarization showed changes of only a few dB (Fig. 3(a)). This means that our EAM was designed to support TE optical mode, and since a high extinction ratio is required for Gaussian-modulated CV-QKD systems, TE polarization should be selected for the proof-of-principle. The current consumption values are shown as dashed curves in Fig. 3(a) and Fig. 3(b), and it can be seen that current consumption in the EAM for TE polarization (from -0.75 mA to -1.03 mA) is higher than when TM polarized light is used (from -0.10 mA to -0.72 mA).

Fig. 4(a) shows the transmission of the top and bottom MZIs (green and yellow lines, respectively) of the IQ modulator (see Fig. 1(a)) when a sweep in current is applied to one of the CIPSs. The MZIs show an ER of 25 and 22 dB for the top and bottom MZI, respectively. The modulation efficiency, defined as $V_{pi} \cdot L$, was close to 1.13 Vmm, considering $V_{pi}$ = 1.3 V and modulators' length of 875 μm. Moreover, the difference in the optical response of the two MZIs and the asymmetry of each of their responses produces a distortion to the optical signals when IQ modulation is performed. Hence, digital compensation of the signals is necessary to correct the optical distortion. To ensure the orthogonality between the two MZIs, we use a TOPS calibrated using a coherent detection system at the output of the chip. A constant DC value is sufficient to maintain the orthogonality between the two components. A phase-shift keying signal with few symbols was used to calculate predistortion coefficients and extrapolate them to the Gaussian modulation. Finally, Fig. 4 (b) presents the electro-optical VOA performance when just a TOPS is used (see Fig. 1(e)). A current sweep from 0 to 42 mA was applied and 33 dB attenuation was obtained at 35 mA. The obtained results show that our PIC TX meets the required ER for the implementation of pulsed GMCS CV-QKD.

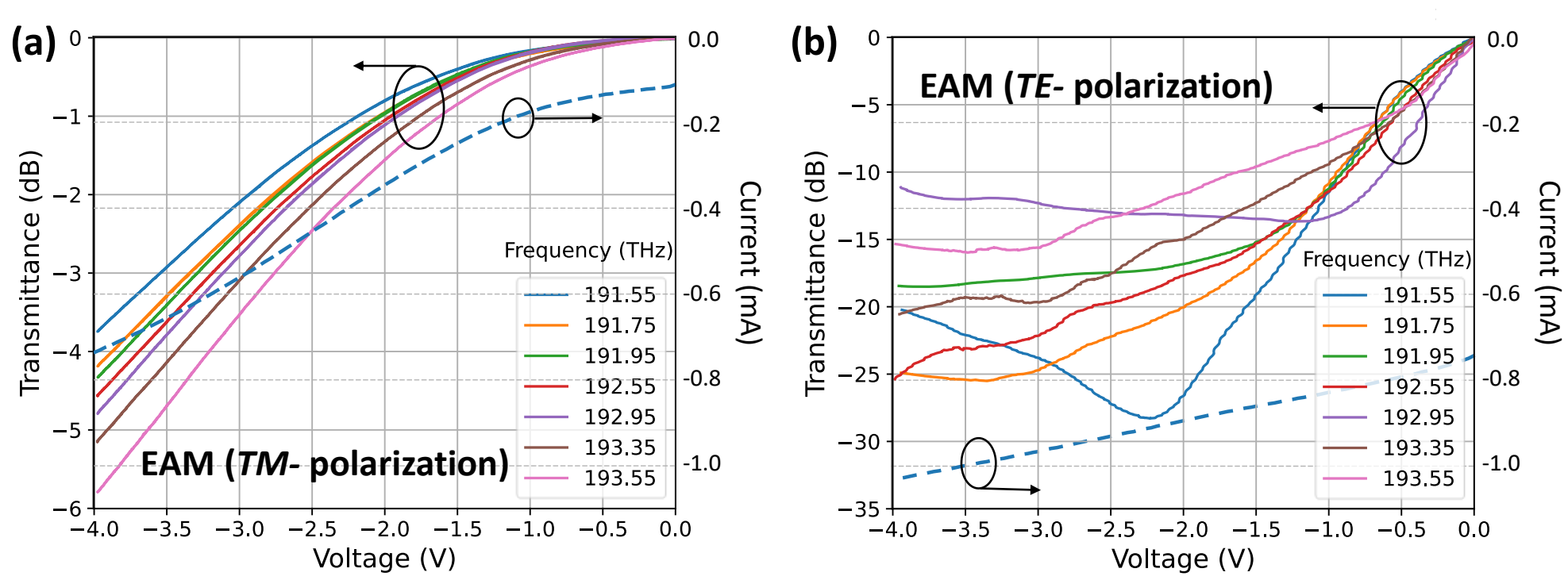

Fig. 3. Transmittance and current as a function of the voltage applied to the EAM for different optical frequencies and two polarizations of the incident light: (a) TM- and (b) TE- polarization. Dashed light blue lines are the current consumption of the EAM.

Regarding IL, losses in the TX are not critical since the light is attenuated to the quantum level. However, sufficient output power is needed for measuring and calculating $V_A$. In our case, the total IL from In IQ to Out IQ (Fig. 1(a)) was 37.9 dB, accounting for fiber coupling, components, and monitoring ports in the PIC TX. The IL for the EAM at TE polarization was 12.4 dB. Combining the EAM and IQ modulator resulted in high IL, preventing power measurement due to the shared LO scheme. To address this, external AM with lower IL (Fig. 2 (a)) was used instead of the EAM, though pulse shaping modulation could eliminate the need for an EAM [29,30].

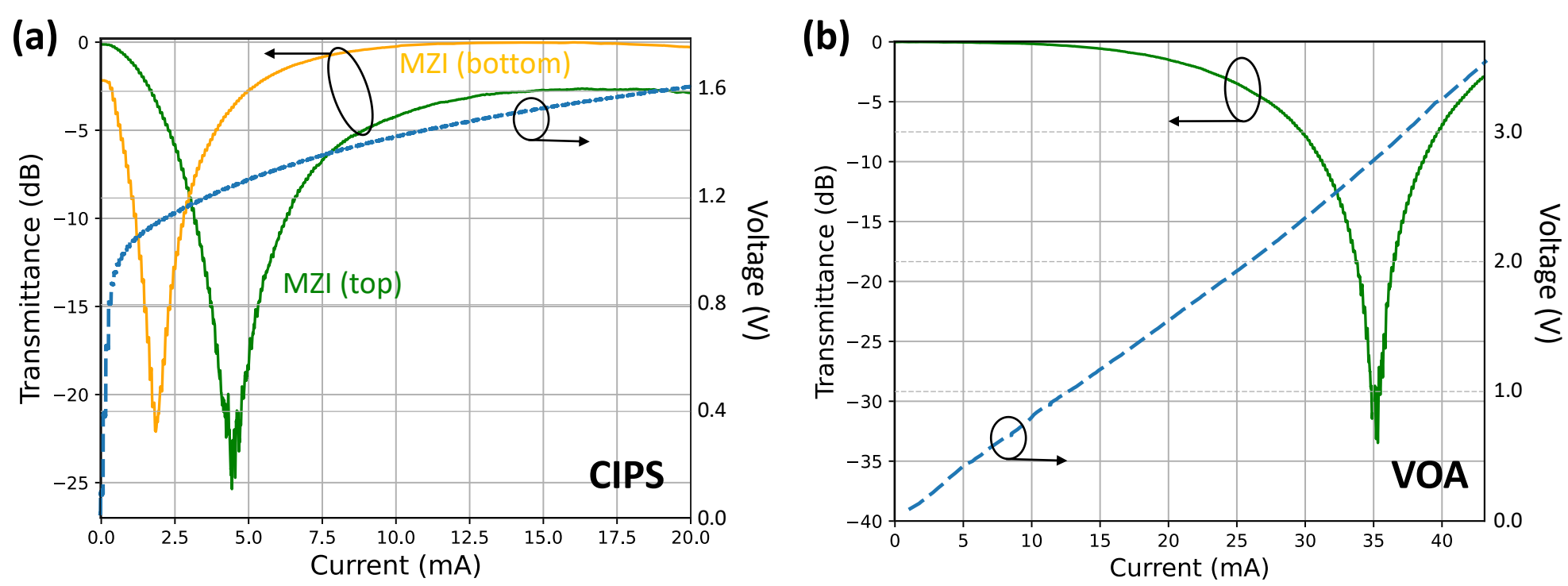

Fig. 4. Electro-optical response of the MZIs when a sweep in current is applied to the (a) CIPS and (b) VOA. Dashed light blue lines are the electronic current consumptions of the components.

### *3.2 CV-QKD experimental results*

The outcomes of the experimental setup described in Section 2.2 were analysed to assess the performance of the characterized PIC in a realistic CV-QKD setup over an 11 km optical fiber channel. To perform this assessment, a series of DSP algorithms were applied to the measured RTO samples to obtain, at Bob's site, a symbol sequence comparable to Alice's. A comparison between a subset of Alice's and Bob's symbols allows the estimation of all the relevant parameters required to bound the information that potential eavesdroppers can have on the shared information between Alice and Bob. This is sufficient to have a reliable estimation of the potential secret key rate that the system can provide under these particular experimental conditions and, up to a certain point, extrapolate its performance at different distances. In practice, to obtain the final key, further steps of error correction and privacy amplification need to be performed, but their implementation lays out of the scope of this project. We chose here to use reconciliation efficiency parameters reported in the literature to take into account the imperfections associated to these steps and estimate the SKR [31].

We generate in the evaluated PIC an optical signal following a 2D-Gaussian distribution in phase-space. To do this, we divide the digital complex sequence into its real and imaginary parts, to which we apply a digital pre-compensation according to the corresponding modulation response (see Fig.4 (a)), and we send these values to the analog electronic output chain (DAC, RF amplifier, RF cable/probe, chip pad). The orthogonality between the two phase-space components is ensured by a TOPS in one of the arms, and the mean output photon value is controlled by the VOA. The different steps of the DSP process are illustrated in Fig. 5 and were previously presented in [26]. The output of each detector is sampled using the RTO, producing two real sequences of samples that can be interpreted as a sequence of complex numbers with a phase-space density as the one shown in Fig. 5(a). The sequences are downsampled to the symbol rate by periodically choosing the samples that maximize the energy of the resulting signal, allowing for the conversion from samples to symbols (Fig. 5(b)). At this stage, the innermost symbols correspond to the quantum information while the outermost ones correspond to the reference symbols. The ring shape in the phase space is due to the phase noise introduced by the 11 km of fiber and the laser (relative phase of the LO and the signal), which needs to be estimated and corrected in the subsequent DSP element. The reference symbols are used to predict the instantaneous phase drift of the signals [5]. The constellation after applying the correction of the estimated phase is shown in Fig. 5(c). After removing the references, we get the constellation shown in Fig. 5(d), where the outer histograms are the distributions of the X and P quadratures with a zero-centered Gaussian distribution. After the phase recovery of the Gaussian symbols received by Bob, we perform pattern synchronization between Alice and Bob using cross-correlation [5], and then we are ready to perform the parameter estimation (PE) of QKD dynamic parameters (excess noise and transmittance of the channel) and calculate the potential SKR. Alice's modulation variance is estimated from the optical power meter at her output. The rest of the parameters are assumed to remain constant and calibrated beforehand.

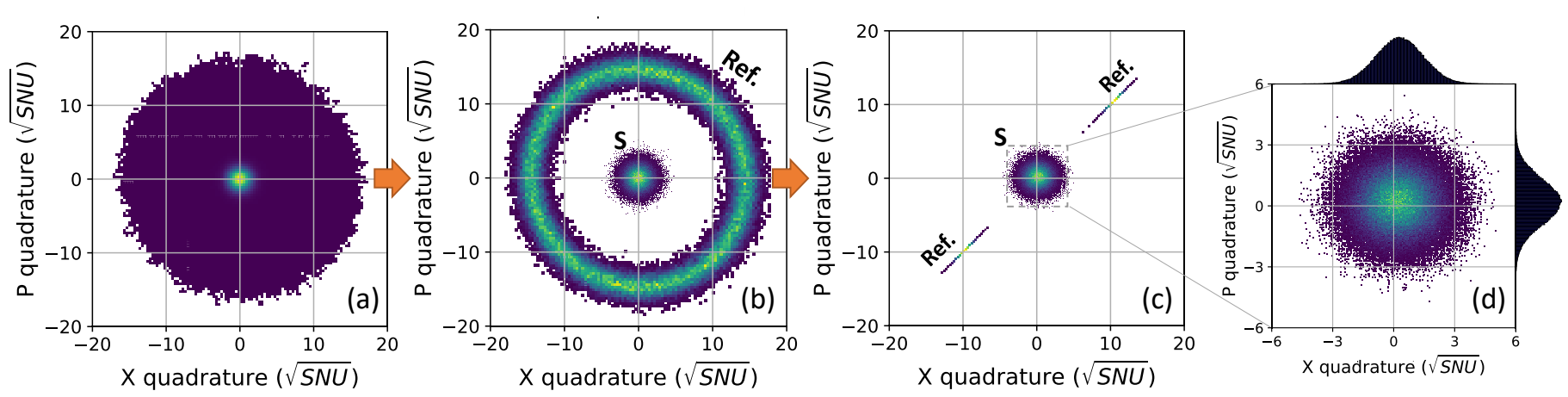

Fig. 5. Received optical constellation after: (a) the acquisition using the oscilloscope; (b) downsampling; and (c) phase recovery of the references (Refs) and symbols (S). (d) Phase-space density of the quantum symbols measured at the coherent receiver output after downsampling and phase recovery for Gaussian signals. (outside) Histogram for the amplitude of X and P quadratures of received optical Gaussian signal.

To this end, we will denote the excess noise estimation at the output of Alice's site as $\xi_A$, following the expression $\xi_A = 2\xi_{B_q}/(\eta T)$, where $\xi_{B_q}$is the excess noise estimation at Bob's site in shot noise units (SNUs) —the total excess noise at Bob's site can be calculated by $\xi_B = 2\xi_{B_q}$—, $\eta$ is the detection efficiency, and $T$ is the transmittance of the channel. The estimated excess noise includes effects that are due to the uncontrolled imperfections in the generation and detection of the states, such as incorrect pre-compensation or biased modulation. This allows the system to be secure, at the cost of performance, even if some modulation issues are not perfectly addressed. As usual, the channel loss is considered a potential security threat, controlled by an eavesdropper. Our PE analysis will focus on the estimation of $\xi_{B_q}$ and $T$, assuming that the other relevant parameters can be estimated with arbitrary precision (shown in Table 1). We can then apply the well-known Devetak-Winter formula to obtain an estimation of the SKR. In the formula below, $\beta$ is the reconciliation efficiency, $V_A$ is the symbol variance at the output of Alice's in SNU, $\nu_{\text{elec}}$ is the electronic noise of the detector in SNU, $R_{\text{eff}}$ is the effective symbol rate of quantum symbols in Baud=symbol/s, $I_{AB}$ is the mutual information between the sender (Alice) and the receiver (Bob) and $\chi_{BE}$ is the Holevo bound (both in bit/symbol):

$$SKR = \left(\frac{N-m}{N}\right)\left[\beta I_{AB}(V_A, \boldsymbol{T}, \boldsymbol{\xi_{B_q}}, \nu_{\text{elec}}, \eta) - \chi_{BE}(V_A, \boldsymbol{T}, \boldsymbol{\xi_{B_q}}, \nu_{\text{elec}}, \eta)\right] R_{\text{eff}}. \qquad (1)$$

The initial quotient indicates the ratio of symbols not used in the PE with respect to the total exchanged symbols. In practice, QKD systems operate during a finite time exchanging a finite number of symbols $N$, of which $m < N$ can be used for the PE, leaving a fraction *(N-m)/N* available for secret key generation. This fact also implies having an imperfect estimate of the parameters. The effects of this finite precision in the PE is still an active research area for the general case of any constellation (in particular, discrete modulation) [2,32–34] while the security proof is complete for the case of Gaussian modulation and heterodyne detection [9]. In the asymptotic regime, where it is assumed that the statistics obtained with the available symbols are equivalent to those that would be obtained from an infinite number of symbols [35,36], we obtained a potential SKR of 156 kbps without restraints on the number of symbols used for parameter estimation. In this case, the measured excess noise was 0.027 SNU. If we assume that one-out-of-two symbols are used for parameter estimation and the rest is destined for generating the key $\left((N-m)/N=1/2\right)$ [37], the achievable SKR in this optimistic asymptotic regime would be 78 kbps.

**Table 1.** Transmission parameters used for asymptotic and finite size approach.

| | **Parameter** | **Symbol** | **Value** |
|---|---|---|---|
| Static parameters | Electronic noise variance | $v_{\text{elec}}$ | 13 mSNU |
| | Reconciliation efficiency | $\beta$ | 0.95 [31] |
| | Detection efficiency | $\eta$ | 0.296 |
| | The ratio of the intensity of reference pulses to quantum pulses | $\rho$ | 342.6 |
| | Effective quantum pulse rate | $R_{\text{eff}}$ | 8 Msymbol/s |
| | Parameter estimation security parameter (finite size regime) | $\epsilon$ | $10^{-10}$ |
| Dynamic parameters | Alice's modulation variance | $V_A$ | 2.778 SNU |
| | Total excess noise measured at Bob (asymptotic regime) | $\xi_B$ | 0.027 SNU |
| | Transmittance of the channel (asymptotic regime) | $T$ | 0.624 |
| | Secret key rate @ 11km ((*N-m*)/*N*=1, asymptotic regime) | *SKR* | 156 kbps |
| | Secret key rate @ 11km ((*N-m*)/*N*=1/2, asymptotic regime) | *SKR* | 78 kbps |

An option to estimate more realistic bounds, taking into account the finite number of elements used in parameter estimation, is to use the proposal of [38] which takes worst-case estimators assuming Gaussian properties in the estimated values. The estimates of $\xi_{B_q}$ and $T$ will be different for the asymptotic and the finite size (FS) case, and a detailed derivation of their calculation is included in the Supplementary material. With this method, although the SKR approaches the asymptotic one as *m* increases, relatively low values of *m* have a serious impact in the SKR and range. In Fig. 6 we show, for the asymptotic and various finite size values, an extrapolation of the results obtained for the 11 km optical fiber channel to other distances, assuming that the estimated excess noise does not differ significantly in that range. Practical limitations in our acquisition setup limited our number of total symbols to $N = 3.08 \times 10^6$ symbols, restricting our communication range in this regime to a few km for a parameter estimation security parameter $\epsilon = 10^{-10}$. This limitation is not related to the PIC, and could be overcome with more suitable acquisition devices. Also, different values of the security parameter could be considered, as discussed in the Supplementary material.

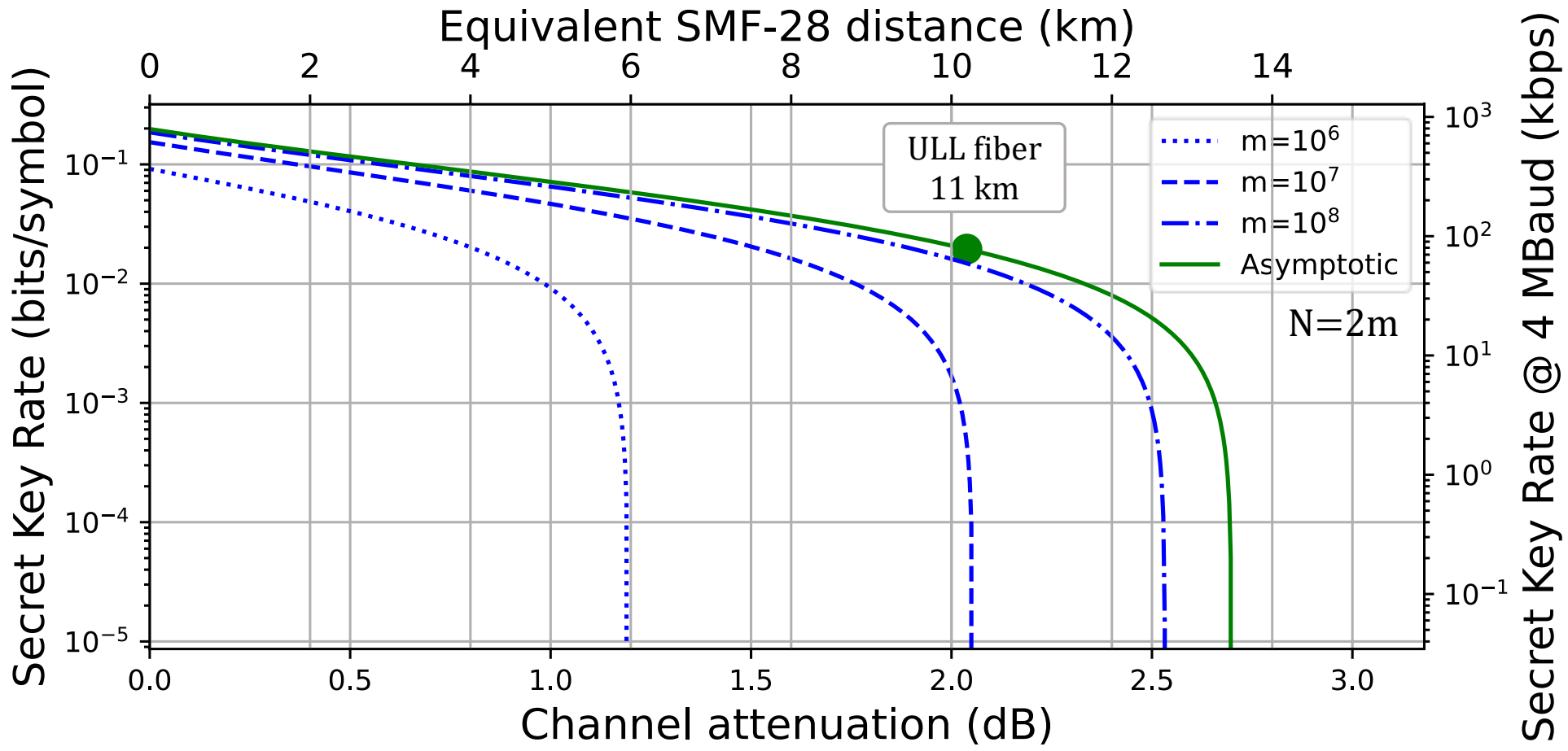

Fig. 6. Comparison of the expected SKR as a function of the channel attenuation using asymptotic (green line) and finite size (dashed lines) analysis calculated at $\beta$=0.95, $\epsilon$=$10^{-10}$, and different values of *m*. The experimental result obtained using an ULL fiber is marked with a green dot.

## 4. Conclusions and outlook

We have reported on the performance of an InP-based PIC operating as a coherent encoder in a realistic CV-QKD setup. In the first part of our study, we showed the design and electro-optical characterization of the PIC TX. The device exhibited good performance in terms of extinction ratio, making it well-suited for the pulsed GMCS CV-QKD protocol, but the flexible characteristics of this research-grade PIC where detrimental in terms of the insertion loss of the entire chain. In the second part, we conducted experimental demonstrations showcasing the viability of our PIC TX in CV-QKD systems. Using a GMCS protocol over a distance of 11 km optical fiber and a TLO scheme, we achieved a potential SKR of 78 kbps in the asymptotic regime. These results are promising and serve as a proof of concept for applying an InP PIC TX in CV-QKD systems. Although they could be improved by increasing, for instance, the symbol repetition rate and expanding the acquisition device memory size to cover finite size regimes, our results open up the possibility for the use of the InP platform for the monolithic integration of CV-QKD systems, offering cost savings, reduced complexity, compact size, and lower power consumption.

Building upon the proposed PIC TX, future advancements in this field could focus on the monolithic integration with other components, such as integrated laser and power meter (photodiode for feedback). Furthermore, there are PIC areas that require improvement, such as reducing losses introduced by certain components (e.g., EAM and IQ modulators), and addressing distortions in IQ modulation. Additionally, pulse shaping modulation should be explored in order to optimize the use of the available dynamic range. Optical and electronic packaging is another relevant factor that would facilitate the testing of these devices. In terms of system-level performance, conducting experiments with more realistic setups (utilizing a real LO) and performing joint demonstrations with an integrated receiver would be important. These improvements would enable the full integration of CV-QKD systems and their widespread use. It is worth noting that weak coherent pulses (WCP) implementations of discrete-variable (DV) QKD protocols would also benefit from the possibility of monolithically integrating lasers and fast modulators offered by InP. In the long term, it would also be possible to integrate transmitters that can support both DV and CV-QKD systems [39].

**Funding.** Agència de Gestió d'Ajuts Universitaris i de Recerca (2021 SGR 01458), H2020 Marie Skłodowska-Curie Actions (713729); Horizon 2020 Framework Programme (CiViQ, 820466); European Union (QSNP, 101114043); Grant PTQ2021-012121 [MCIN/AEI/10.13039/50110001103], DGR-Next Generation Catalonia.

**Acknowledgments.** This project is funded by the Departament de Recerca i Universitats de la Generalitat de Catalunya (2021 SGR 01458). This work was partially funded by CEX2019- 000910-S (MCIN/AEI/10.13039/501100011033), Fundació Cellex, Fundació Mir-Puig, and Generalitat de Catalunya through CERCA. This study was supported by MCIN with funding from European Union NextGenerationEU(PRTR-C17.I1) and by Generalitat de Catalunya

**Disclosures.** The authors declare no conflicts of interest.

**Data availability.** Data underlying the results presented in this paper are not publicly available at this time but may be obtained from the authors upon reasonable request.

**Supplemental document.** See Supplement 1 for supporting content.

**SUPPLEMENTAL DOCUMENT**

## 1. Parameter estimation in the asymptotic and finite size scenarios

For protocols with Gaussian modulation and heterodyne detection, the covariance matrix between Alice's mode and one of Bob's two modes ($\Sigma_{AB_{1,2}}$) is expressed, with the variances in shot noise units (SNU), as [1]:

$$\Sigma_{AB_{1,2}} = \begin{pmatrix} (V_A+1)\mathbb{1}_2 & \pm\sqrt{\frac{\eta T}{2}}(V_A^2+2V_A)\sigma_z \\ \pm\sqrt{\frac{\eta T}{2}}(V_A^2+2V_A)\sigma_z & \left(\frac{1}{2}\eta T V_A + \xi_{B_q} + \nu_{\text{elec}} + 1\right)\mathbb{1}_2 \end{pmatrix}, \quad \text{(S1)}$$

where $\eta$ is the detection efficiency, $T$ is the transmittance of the channel, $\nu_{\text{elec}}$ is the electronic noise variance, and $\xi_{B_q}$ is the excess noise variance measured in one of the quadratures. The variance in the quadrature distribution at the transmitter (Alice) $V_A$ and receiver's site (Bob) $V_B$, in the case of heterodyne detection are related as:

$$V_B = \frac{1}{2}\eta T V_A + \xi_{B_q} + \nu_{\text{elec}} + 1, \quad \text{(S2)}$$

and the variance in Bob's measurement conditioned to Alice's data is [1]:

$$V_{B|A} = \xi_{B_q} + \nu_{\text{elec}} + 1, \quad \text{(S3)}$$

which is equivalent to:

$$V_{B|A} = var\left(\sqrt{\frac{\eta T}{2}}\hat{q}_A - \hat{q}_B\right), \quad \text{(S4)}$$

where $\hat{q}_A = \{X_{A_i}, P_{A_i}\}$ and $\hat{q}_B = \{X_{B_i}, P_{B_i}\}$. The variances and different noises are normalized to the SNU.

For practical parameter estimation (PE), the excess noise at Bob's site can be estimated from Eq.(S3) and the conditional variance $V_{B|A}$ in Eq. (S4) [1] as:

$$\xi_{B_q} = var\left(\sqrt{\frac{\eta T}{2}}\hat{q}_A - \hat{q}_B\right) - \nu_{\text{elec}} - 1. \quad \text{(S5)}$$

Hence, the excess noise at Alice's site will be given by, $\xi_A = 2\xi_{B_q}/(\eta T)$. Here the channel loss was considered as a potential security threat controlled by an eavesdropper. A detailed description of the calculation of the rest of the parameters of the QKD system in the asymptotic regime ($T$, $V_A$) can be found in [1,2].

For the finite size analysis, *m* values are used for the PE from the total number of symbols *N* exchanged by Alice and Bob during the protocol. The remaining values ($n=N-m$) are not revealed and can be used for the generation of the secret key. In this scenario, for the PE, let's first consider the lineal model to relate Alice's and Bob's data (*x* and *y*, respectively):

$$y = tx + z, \quad \text{(S6)}$$

with the parametrization $t = \sqrt{\eta T/2}$. From this, we get (when compared with Eq. (S2)) the variance of *x* as $V_A$, variance of *y* as $V_B$, and variance of *z* as:

$$\sigma^2 = \xi_{B_q} + \nu_{\text{elec}} + 1. \quad \text{(S7)}$$

Consequently, from Eq. (S1), the covariance matrix that minimizes the secret key rate (SKR) due to the PE will be expressed as [3]:

$$\Sigma_{AB_{1,2}}^{PE} = \begin{pmatrix} (V_A+1)\mathbb{1}_2 & \pm t_{\min}\sqrt{(V_A^2+2V_A)}\sigma_z \\ \pm t_{\min}\sqrt{(V_A^2+2V_A)}\sigma_z & (t_{\min}^2 V_A + \sigma_{\max}^2)\mathbb{1}_2 \end{pmatrix}, \tag{S8}$$

where $t_{\min}$ is the minimum value of $t$ and $\sigma_{\max}^2$ is the maximum value of $\sigma^2$ that minimize the SKR (these values are conditioned on being compatible with the measured data). This means that we have now bounded our confidence region. Therefore, the PE, taking into consideration the finite size effects, can be calculated from [3]:

$$t_{\min} \approx \hat{t} - \Delta\hat{t} = \hat{t} - z_{\epsilon_{PE}/2}\sqrt{\frac{\hat{\sigma}^2}{mV_A}},$$
$$\sigma_{\max}^2 \approx \hat{\sigma}^2 + \Delta\hat{\sigma}^2 = \hat{\sigma}^2 + z_{\epsilon_{PE}/2}\frac{\hat{\sigma}^2\sqrt{2}}{\sqrt{m}}, \tag{S9}$$

where $\hat{t}$ and $\hat{\sigma}^2$ are the values estimated from the experimental realization, and $z_{\epsilon_{PE}/2} = \sqrt{2}erf^{-1}(1-\epsilon_{PE}) \approx 6.5$ if we consider a security value $\epsilon_{PE} = 10^{-10}$; $erf$ is the error function.

We can rewrite Eq. (S7) as a function of the total excess noise measured at Bob's site:

$$\sigma^2 = \xi_{B_q} + \sigma_0^2$$
$$\xi_{B_q} = (\sigma^2 - \sigma_0^2), \tag{S10}$$

where $\sigma_0^2 = (\nu_{\text{elec}} + 1)$.

Taking into consideration the calibration imperfections of the electronic noise and the shot noise, the estimated excess noise in Eq. (S10) is bounded and re-defined as [4]:

$$\xi_{B_q}^{FS} = [(\hat{\sigma}^2 + \Delta\hat{\sigma}^2) - (\hat{\sigma}_0^2 - \Delta\hat{\sigma}_0^2)], \tag{S11}$$

where $\Delta\hat{\sigma}_0^2$ is similar to $\Delta\hat{\sigma}^2$ in Eq. (S9) and $m'$ is the number of symbols used for the calibration estimation, then:

$$\Delta\hat{\sigma}_0^2 = z_{\epsilon_{PE}/2}\frac{\hat{\sigma}_0^2\sqrt{2}}{\sqrt{m'}}. \tag{S12}$$

## 2. *Secret key estimation in the asymptotic and finite size scenarios*

Following the PE, the SKR in bits per second (bps), taking into consideration the finite-size effects can be computed using the Devetak-Winter formula:

$$SKR_{FS} = \left(\frac{N-m}{N}\right)\Big[\beta I_{AB}\left(V_A, T_{\min}, \xi_{B_q}^{FS}, \nu_{\text{elec}}, \eta\right) - \chi_{BE}\left(V_A, T_{\min}, \xi_{B_q}^{FS}, \nu_{\text{elec}}, \eta\right) - \Delta(n)\Big] R_{\text{eff}}, \tag{S13}$$

where the minimum transmittance of the channel $T_{\min}$ is related to $t_{\min}$ (defined in Eq. (S9)) as $t_{\min} = \sqrt{\eta T_{\min}/2}$. Eq. (S13) is for the case of reverse reconciliation where Alice's information is corrected following Bob's measurements [5], where $\beta$ stands for the reconciliation efficiency (efficiency of the error correcting code), $I_{AB}$ is the mutual information between the sender (Alice) and the receiver (Bob), $\chi_{BE}$ is the Holevo bound, $R_{\text{eff}}$ is the effective quantum pulse rate in Hz or pulses per second and $\Delta(n)$ is the parameter related to the security of the privacy amplification [3]. In this case, it was considered to be zero, because this work was focused on the PE finite size effect.

In the asymptotic limit, it is assumed that the data used for the PE is large enough to make the correction terms $(\Delta\hat{t}, \Delta\hat{\sigma}^2, \Delta\hat{\sigma}_0^2)$ all go to zero [3]. Subsequently, Eq. (S13) in the asymptotic regime is simplified as Eq. (S14):

$$SKR = \left(\frac{N-m}{N}\right)\left[\beta I_{AB}(V_A, T, \xi_{B_q}, \nu_{\text{elec}}, \eta) - \chi_{BE}(V_A, T, \xi_{B_q}, \nu_{\text{elec}}, \eta)\right] R_{\text{eff}}\,. \qquad \text{(S14)}$$

The factor *(N-m)/N* introduces a trade-off between secret key rate and accuracy of the parameter estimation in the finite-size regime. Due to practical constraints in our measurement setup that are not related to the PIC, we are limited to $N = 3.08 \times 10^6$ symbols, which restricts the values of $m$ to the shaded areas of Fig. S1, calculated for a parameter estimation security parameter $\epsilon = 10^{-10}$. These estimates of excess noise and transmittance lead to poor SKR and range using the typical values for reconciliation efficiency ($\beta$ = 0.95) [6], as can be seen in the Fig. 6 of the article (the curve for $m = 10^6$ can be taken as a reference). The finite size effects can be improved by increasing the number of processed symbols, something that is not related to the PIC under test, only the surrounding setup. For this reason, we provide only asymptotic estimates in the article, but we do consider that a fraction of the symbols is revealed to perform parameter estimation.

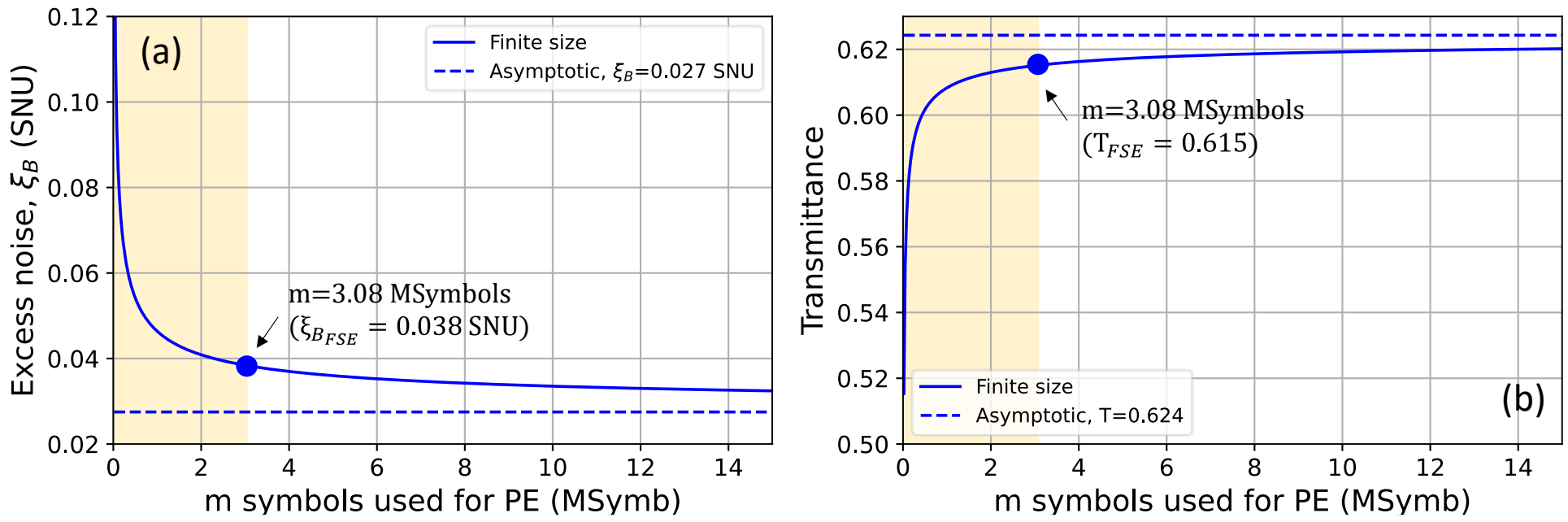

Fig. S1. (a) Total excess noise $\xi_B = 2\xi_{B_q}$ and (b) transmittance of the channel $T$ as a function of the number of symbols used for its estimation using finite size analysis (blue line) and asymptotic analysis (dashed blue line). The blue dots correspond to the finite-size worst-case estimators when $N = m$; in practice, they will be worse. The parameter estimation security parameter considered is $\epsilon = 10^{-10}$.